\documentclass[aps,12pt,notitlepage,final,oneside,onecolumn,nobibnotes,
nofootinbib,superscriptaddress,noshowpacs,centertags,amsmath,amssymb,a4paper]{revtex4}

\usepackage{geometry}
\usepackage{mathtools}

\newcommand{\bs}{\bm \sigma}

\newcommand{\bS}{\bm \Sigma}
\newcommand{\bg}{\bm \gamma}
\newcommand{\bxi}{\bm \xi}
\newcommand{\gz}{\gamma^{0}}
\newcommand{\gf}{\gamma^{5}}

\RequirePackage{graphicx}

\begin{document}

\title{Neutrino propagation in media and  axis of complete polarization}

\author{A.E. Kaloshin}
\email{alexander.e.kaloshin@gmail.com}
\affiliation{Irkutsk State University, K.  Marx str., 1, 664003, Irkutsk, Russia}
\affiliation{Joint Institute for Nuclear Research, 141980, Dubna, Russia}

\author{D.M.Voronin}
\email{dmitry.m.voronin@gmail.com}
\affiliation{Irkutsk State University, K.  Marx str., 1, 664003, Irkutsk, Russia}

\begin{abstract}
We construct a spectral representation of neutrino propagator in moving matter or in external magnetic field. In both cases there exists fixed four-dimensional axis of polarization, such that the corresponding spin projectors commute with propagator. As a result, all eigenvalues of propagator and, consequently,  dispersion laws for neutrino in media are classified according to spin projection onto this axis.
Use of the found spin projectors simplifies essentially the eigenvalue problem and allows to build spectral representation of propagator in moving matter or external magnetic field in analogy with the vacuum propagator.
\end{abstract}

\maketitle

\section{Introduction}
\label{intro}
Neutrino physics,  actively developing in last decades, is related with a wide spectrum of physical problems, including the astrophysical ones. Propagation of neutrinos in a dense  matter or magnetic field leads to modification of neutrinos oscillation picture and appearance of new effects.
The most prominant effect in neutrinos passing through matter is related with resonance amplification of oscillations  (MSW effect) \cite{q16,m5656}, which solves  the solar neutrino  problem, see reviews \cite{q8}-\cite{Bil18}. 

After it  a huge interest was generated  to further investigations of different properties of media and its influence on fine aspects of flavor oscillations. Different methods were developed for solution of corresponding equations in media with varied density \cite{Haxton:1986dm}-\cite{Balantekin:1997jp}. 
The movement of matter and its polarization, which can arise in magnetic field were taken into account \cite{Nunokawa:1997dp}-\cite{EstebanPretel:2008ni}. Interesting results were obtained in investigations of spin dynamics in matter and transitions between different spin states \cite{Lobanov:2001ar}, \cite{Dvornikov:2002rs},  \cite{Lobanov:2002ur}.  As for applications in astrophysics, there exists a variety of conditions  for neutrinos propagation: in Earth, Sun, in vicinity of Supernova -- see reveiws \cite{Duan:2009cd}, \cite{Volpe:2016bkp}. We mentioned here only some aspects of investigations and only small part of  relevent publications.

Possibility for neutrino to have an anomalious magnetic moment  and its experimental manifestations was discussed for a long time \cite{Cis71}-\cite{Ego00}.													
In Standard Model (SM) the magnetic moment of neutrino is arised due to loop corrections and is proportional to neutrino mass. It leads to extreme smallness of neutrino magnetic moment in SM, so the present-day interest for this subject (both in theory and experiment)  is related first of all with search of new physics beyond the SM \cite{Zhi11,Giu15}.

Most transparent way to describe mixing and oscillations phenomena in neutrinos system is to use the quantum-mechanical equations (Schr\"odinger or Dirac), but the most justified is the Quantum Field Theory (QFT) approach, where production, propagation and detection of neutrino looks like a macroscopic Feynman diagram\cite{Gri96}-\cite{Dvo11}.  The necessary element of such description is the neutrino propagator. 

In the present paper we build a spectral representation of neutrino propagator in matter moving with constant velocity  or in constant homogenious magnetic field \footnote{Short version of this paper without discussion of propagation in external magnetic field was published in \cite{VK}.}. In this representation based on the eigenvalue  problem a propagator looks as a sum of single poles, accompanied by orthogonal matrix projectors. Such form of propagator gives the simplest and most convenient algebraic construction and means in fact a complete diagonalization. A spectral representation was discussed earlier for dressed fermion propagator in theory with parity violation \cite{q12} and for dressed matrix propagator with mixing of few fermionic fields \cite{q98}.
Note that the problem of the neutrino propagator in the presence of matter,
including account for possible effect of matter motion,
was also discussed in \cite{Pivovarov:2005cu} (see also \cite{Studenikin:2008qk}). 

In solving the eigenvalue problem for neutrino propagator in media, we find a new aspect, related with polarization of neutrino: there exist  spin projectors  with fixed polarization 4-vector, commuting with propagator.  
The properties of these spin projectors (both in matter and magnetic field) allows to reduce the algebraic problem for media to the vacuum case,
properties  of media modify only scalar coefficients of matrix equations. 

In Section 2 we construct the spectral representation of neutrino propagator in a matter, moving with constant velocity. The key moment is the presence of generalized spin projectors  (\ref{Sigma}), commuting with propagator, which allows to simplify the eigenstate problem  (\ref{eq:5}) and to get answer for inverse propagator of the most general view (\ref{formula2}). We discuss also the particular case of propagator in framework of Standard Model, in this case it is easy to see that the spin projection on the complete polarization axis $z^\mu$ is not conserved.

In Section 3  the spectral representation is build for neutrino propagator in a constant magnetic field. In this case there exists the fixed axis of complete polarization $z^\mu$ and again the corresponding spin projector $\Sigma(z)$ commutes with propagator. This property allows to use the same trick (reducing of number of $\gamma$-matrix structures ``under observation'' of spin projectors) to obtain an analytical expressions for eigenvalues and eigenprojectors. 

In Appendix the main facts on spectral representation for matrix of general form and some details of this representation for fermion propagator in vacuum and media are collected.

\section{Propagator in moving matter and spin projectors}
\label{sec:nonrel}
\subsection{Axis of complete polarization and basis}
When considering a neutrino oscillations in frameworks of quantum field theory, the central object is the neutrino propagator. In media there exist two 4-dimensional vectors: momentum of particle $p$  and matter velocity $u$, so altogether there exist eight $\gamma$-matrix structures in decomposition of propagator, if parity is not conserved. Most general expression for inverse propagator can be written as
\begin{eqnarray}\label{formula2}
S(p,u)=G^{-1} =s_{1}I+s_{2}\hat{p}+s_{3}\hat{u}+s_{4}\sigma^{\mu\nu}p_{\mu}u_{\nu}+\nonumber\\+s_{5}i\varepsilon^{\mu\nu\lambda\rho}\sigma^{\mu\nu}u_{\lambda}p_{\rho}+s_{6}\gamma^{5}+s_{7}\hat{p}\gamma^{5}+s_{8}\hat{u}\gamma^{5},
\end{eqnarray}
where $s_{i}$ are scalar function dependent on invariants. 

Below we will solve the eigenvalue problem for inverse propagator of general form. As a starting point  it is convenient to introduce $\gamma$-matrix basis with simple multiplicative properties. 

Firts of all, let us introduce the four-vector  $z^{\mu}$, which is a linear combination of two vectors $p$,  $u$ and has properties of fermion polarization vector \footnote{This vector was used earlier \cite{q11} for some algebraic simplification of propagator in matter.}:
\begin{equation}\label{123sa}  
z^{\mu}p_{\mu}=0,  \ \ \ z^{2}=-1. 
\end{equation} 

Orthogonal to momentum combination is
\begin{equation}\label{vec_z}
z^{\mu}=b\ (p^{\mu}(up)-u^{\mu}p^{2}),
\end{equation}
where $b$ is the normalization factor, $b=[p^{2}((up)^{2}-p^{2})]^{-1/2}$.

To clarify its properties, let us consider firstly the rest matter $({\bf{u}}=0,u^{0}=1)$. In this case 
\begin{equation}
z^{\mu}=b({\bf{p}}^{2},p^{0}{\bf{p}})
\end{equation}
and its square is
\begin{equation}
z_{\mu}z^{\mu}=-b^{2}{\bf{p}}^{2}(p_{\nu}p^{\nu}).
\end{equation}

Thus we see, that for time-like   momentum $p^{\mu}$ the vector $z^{\mu}$ is space-like one
\footnote{Note that for space-like momentun $p^{2}<0$ the polarization vector $z^{\mu}$ becomes imaginary. But the product  $\hat{z}\hat{n}=\hat{z}\hat{p}/W$ in spin projector (\ref{Sigma}) remains real.}. 

Then, having the vector $z$, one can construct the generalized off-shell spin projectors
\footnote{We call them as generalized because of  appearence of additional factor $\hat{n}$. But in fact the Eq. (\ref{Sigma}) is the most general form of spin projectors at dressing of fermion propagator in theories with parity violation --- see details in \cite{q98}. The same modification (\ref{Sigma}) of a naive spin projector arises in matter -- may be not accidently. }:
\begin{equation}\label{Sigma}
\Sigma^{\pm}=\frac{1}{2}(1\pm\gamma^{5}\hat{z}\hat{n}),~~~ \Sigma^{\pm}\Sigma^{\pm}=\Sigma^{\pm},~~~\Sigma^{\pm}\Sigma^{\mp}=0,
\end{equation}
where $n^{\mu}=p^{\mu}/W,\ \  W=\sqrt{p^{2}}$. The appeared matrix $\gamma^{5}\hat{z}\hat{n}$ may be rewritten as
\begin{equation}\label{}
\gamma^{5}\hat{z}\hat{n}=\gamma^{5} \sigma^{\alpha\beta}z_{\alpha}n_{\beta}=-bW\gf\sigma^{\alpha\beta}u_{\alpha}p_{\beta},\ \ \ \ \ \sigma^{\alpha\beta}=\frac{1}{2}[\gamma^{\alpha}, \gamma^{\beta}].
\end{equation}

After it one can see that $\Sigma^{\pm}$ commute with all $\gamma$-matrices in decomposition of inverse propagator (\ref{formula2}). Multiplying the inverse propagator $S(p,u)$ (\ref{formula2}) by unit matrix
\begin{equation}
S=(\Sigma^{+}(z)+\Sigma^{-}(z))S \equiv S^+ + S^- ,
\end{equation}
one obtains two orthogonal contributions  $S^+, S^-$.

One more useful property of $\Sigma^{\pm}$ is that ``under observation'' of the spin projectors (i.e. in $S^+, S^-$ terms) $\gamma$-matrix structures may be simplified. Namely: \\ $\gamma$-matrices, which contain the matter velocity $u^{\mu}$ may be transformed to the set of four matrices without velocity:  $I, \hat{p}, \gamma^{5}, \hat{p}\gamma^{5}$. For example, one can rewrite the term $\hat{u}$ in (\ref{formula2}) as a linear combination $\hat{p}$ and $\hat{z}$ and to use the projector property $(\Sigma^{+}\cdot\gf\hat{z}\hat{n}=\Sigma^{+})$:
\begin{equation}
\Sigma^{+}\hat{u}=\Sigma^{+}(a_1\hat{p}+a_2\hat{z})=\Sigma^{+}(z)(a_1\hat{p}-\frac{a_2}{W}\hat{p}\gamma^{5}).
\end{equation}

After this simplification we have the vacuum set of Dirac matrices $I, \hat{p}, \gamma^{5}, \hat{p}\gamma^{5}$ and it's convenient to introduce the off-shell momentum projections:
\begin{equation}
\Lambda^{\pm}=\frac{1}{2}(1\pm\hat{n}),~~~n^{\mu}=\frac{p^{\mu}}{W}
\end{equation}
orthogonal to each other.

Having the momentum $\Lambda^{\pm}$ and spin projectors  $\Sigma^{\pm}$, one can build the basis, which will be used below in the eigenvalue problem 
\begin{eqnarray}\label{1}
R_{1}=\Sigma^{-}\Lambda^{+},~~~~~~R_{5}=\Sigma^{+}\Lambda^{+},~~~\nonumber\\
R_{2}=\Sigma^{-}\Lambda^{-},~~~~~~R_{6}=\Sigma^{+}\Lambda^{-},~~~\nonumber\\
R_{3}=\Sigma^{-}\Lambda^{+}\gamma^{5},~~~R_{7}=\Sigma^{+}\Lambda^{+}\gamma^{5},\nonumber\\
R_{4}=\Sigma^{-}\Lambda^{-}\gamma^{5},~~~R_{8}=\Sigma^{+}\Lambda^{-}\gamma^{5}.
\end{eqnarray} 
Multiplicative properties of the basis (\ref{1}) are presented in Table 1, where column elements   multiply from left  the row elements.

The inverse propagator (\ref{formula2}) may be written as decomposition in this basis \begin{equation}\label{decomp}
S(p,u)= \sum_{i=1}^{4} R_i S_i (p^2,pu) + \sum_{i=5}^{8} R_i S_i (p^2,pu),
\end{equation}
where these two sums are orthogonal to each other.

\begin{table}[h]
	\caption{Multiplicative properties of the matrix basis (\ref{1})}
	\label{lklhjlksd}
	\begin{center}
		\begin{tabular}{|c|c|c|c|c||c|c|c|c|c|} \hline      
			&$R_{1}$&$R_{2}$&$R_{3}$&$R_{4}$&$R_{5}$&$R_{6}$&$R_{7}$&$R_{8}$ \\ \hline
			$R_{1}$&$R_{1}$&0&$R_{3}$&0&0&0&0&0\\ 
			$R_{2}$&0&$R_{2}$&0&$R_{4}$&0&0&0&0\\   
			$R_{3}$&0&$R_{3}$&0&$R_{1}$&0&0&0&0\\ 
			$R_{4}$&$R_{4}$&0&$R_{2}$&0&0&0&0&0\\ \hline\hline
			$R_{5}$&0&0&0&0&$R_{5}$&0&$R_{7}$&0\\ 
			$R_{6}$&0&0&0&0&0&$R_{6}$&0&$R_{8}$\\ 
			$R_{7}$&0&0&0&0&0&$R_{7}$&0&$R_{5}$\\ 
			$R_{8}$&0&0&0&0&$R_{8}$&0&$R_{6}$&0\\ \hline
		\end{tabular}
	\end{center}
\end{table}

It can be seen from Table 1, that with use of the basis  (\ref{1}) the eigenvalue problem for inverse propagator (\ref{decomp}) is separeted into two different problems: one for $R_{1}..R_{4}$ and another for  $R_{5}..R_{8}$. Every problem has two different eigenvalues.

\subsection{Spectral representation of propagator in matter} 
\label{sec:rel}

Let us recall that in quantum mechanics the term spectral representation of linear hermitian operator $\hat{A}$ means the following representation \cite{q10}
\begin{equation}
\hat{A}=\sum\lambda_{i}|i\rangle\langle i| =\sum\lambda_{i} \Pi_{i},
\end{equation}	
which contains the eigenvalues $\lambda_{i}$ and eigenprojectors $\Pi_{i}=|i\rangle\langle i|$.
\begin{equation}
\hat{A}|i\rangle=\lambda_{i}|i\rangle.
\end{equation}

Orthonornality of the vector system leads to property of orthogonality of projectors
\begin{equation}
\Pi_{i}\Pi_{k}=\delta_{ik}\Pi_{k} .
\end{equation} 
If an operator is not hermitian, to build a spectral representation one needs to solve two eigenvalue problems: left and right ones (see details in \ref{sec:eigen}).

We want to construct a spectral representation for inverse propagator of general form  (\ref{formula2}), (\ref{decomp}), so we should solve the eigenvalue problem 		
\begin{equation}\label{eq:5}
S\Pi_i = \lambda_i \Pi_i .
\end{equation}

Note that we are solving  eigenvalue problem in a matrix form, i.e. from the begining we are looking for eigenprojectors $\Pi_i$ instead of eigenvectors. It can be done with use of $\gamma$-matrix basis and it allows to avoid cumbersome intermediate formulae. As for non-hermi-\\tiance of fermion propagator: it is enough to solve the left problem and to require the orthogonality of projectors, see \cite{q12}.   

When we solve for this problem, we get the spectral representation of inverse propagator in moving matter:
\begin{equation}\label{RS}
S(p,u)=\sum_{i=1}^{4} \lambda_{i}\Pi_{i}. 
\end{equation}

If the eigenprojectors set is the complete orthogonal system, then propagator is easily obtained by reversing of (\ref{RS}) and looks very simple
\begin{equation}\label{spec_G}
G(p,u)=\sum_{i=1}^{4} \frac{1}{\lambda_{i}}\Pi_{i} ,
\end{equation}
as a sum of single poles, accompanied by corresponding orthogonal projectors.

The use of the matrix basis 	(\ref{1}) simplifies essentially solution of eigenvalue problem. The eigenprojects also may be found in form of decomposition in this basis, orthogonality of spin projectors (see Table 1) leads to more simple problems, where only first or second quartet in (\ref{1}) is involved. It was noted in above that ``under observation'' of the spin projectors all gamma-matrices in (\ref{formula2})  turn into the set $I, \hat{p}, \gamma^{5}, \hat{p}\gamma^{5}$. So, for example, the eigenstate problem for first quartet  of basis elements
\begin{equation}\label{RS1}
\left( \sum_{k=1}^{4}R_{k}  S_{k} \right)\cdot \left( \sum_{i=1}^{4}R_{i}  A_{i} \right)  = \lambda \left( \sum_{i=1}^{4}R_{i}  A_{i} \right)   
\end{equation}
coinsides in fact with the 
eigenstate problem for dressed vacuum propagator with parity violation \cite{q12}. The presence of matter leads  only to appearence of spin projector in (\ref{1}) (do not changing an algebra) and modification of scalar coefficients. Besides, as compared with the vacuum case, there appears twice as much eigenvalues, which arise from two different square equations.

Repeating the algebraic operations from \cite{q12}, one can write an answer for eigenvalue problem in most general case. Eigenvalues and eigenprojectors are looking as:
\begin{eqnarray}
\lambda_{1,2}=\frac{S_{1}+S_{2}}{2}\pm\sqrt{\Big(\frac{S_{1}-S_{2}}{2}\Big)^{2} + S_{3}S_{4}}~, \nonumber\\
\label{eig_1}	
\Pi_{1}=\frac{1}{\lambda_{2}-\lambda_{1}}\Big((S_{2}-\lambda_{1})R_{1}+(S_{1}-
\lambda_{1})R_{2}-\nonumber\\-S_{3}R_{3}-S_{4}R_{4}\Big), \\	
\Pi_{2}=\frac{1}{\lambda_{1}-\lambda_{2}}\Big((S_{2}-\lambda_{2})R_{1}+(S_{1}-
\lambda_{2})R_{2}-
\nonumber\\-S_{3}R_{3}-S_{4}R_{4}\Big),\nonumber \\		\lambda_{3,4}=\frac{S_{5}+S_{6}}{2}\pm\sqrt{\Big(\frac{S_{5}-S_{6}}{2}\Big)^{2} + S_{7}S_{8}}~, \nonumber\\
\label{eig_2}
\Pi_{3}=\frac{1}{\lambda_{4}-\lambda_{3}}\Big((S_{6}-\lambda_{3})R_{5}+(S_{5}-
\lambda_{3})R_{6}-\nonumber\\-S_{7}R_{7}-S_{8}R_{8}\Big), \\
\Pi_{4}=\frac{1}{\lambda_{3}-\lambda_{4}}\Big((S_{6}-\lambda_{4})R_{5}+(S_{5}-
\lambda_{4})R_{6}-\nonumber\\-S_{7}R_{7}-S_{8}R_{8}\Big)\nonumber.
\end{eqnarray}

Here $S_{i}$ are the coefficients of decomposition of inverse propagator in the basis (\ref{decomp}). Recall that the indexies $1,2$ refers to $S^{-}$ (i.e to first quartet in (\ref{decomp}), and $3,4$ to contribution $S^{+}$.

The obtained eigenprojectors have the following properties:
\begin{enumerate}
	\item$S\Pi_{k}=\lambda_{k}\Pi_{k}$, \ \ \ \ \ \ k=1 \dots 4 \ ,
	\item$\Pi_{i}\Pi_{j}=\delta_{ij}\Pi_{j}$,
	\item$\sum\limits_{i=1}^4\Pi_{i}=1$.
\end{enumerate}

The introduced by us four-vector $z^{\mu}$ (\ref{vec_z}) plays role of the complete polarization axis and all eigenvalues are classified by the projection of spin onto this axis. In contrast to vacuum, this axis is not arbitrary -- see some details in \ref{sec:prop}. As it will be seen from discussion of Standard Model case, the projection on this axis is not conserved in general case.

\subsection{Standard Model propagator} 

In the case of SM a fermion propagator in matter looks like: 
\begin{equation}\label{dddf}
S(p,u) = \hat{p} - m-\alpha\hat{u}(1-\gamma^{5}),
\end{equation}
where $\alpha$ is some constant dependent on properties of media and flavour.

Note, that  $\alpha$ and mass term are in fact matrices $n \times n$ because of mixing of neutrinos. But here we consider propagation of neutrino of one type, so in this approximation the propagator will contain only a diagonal element of a flavour matrix. For example, in case of electron neutrino \cite{Kuo89} 
\begin{equation*}
\alpha^{(\nu_{e})}= \frac{G_{F}}{\sqrt{2}}(n_{e}(1+4\sin^{2}\theta_{W})+n_{p}(1-4\sin^{2}\theta_{W}) - n_{n}),
\end{equation*}
where $n_{e}, n_{p}, n_{n}$ are densities of matter particles, $\theta_{W}$ is the Weinberg  angle. 

Let us write down the coefficients of decomposition in two bases:  $\gamma$-matrix (\ref{formula2})  and  $R$-basis   (\ref{decomp}):
\begin{eqnarray}\label{vvbnmkl}
s_{1}=-m,~~~S_{1}=-m+W(1+K^{+});\nonumber\\
s_{2}=1,~~~~~~~S_{2}=-m-W(1+K^{+});\nonumber\\
s_{3}=\alpha,~~~~~~~S_{3}=-WK^{+};~~~~~~~~~~~~~~\nonumber\\
s_{4}=0,~~~~~~~~S_{4}=WK^{+};~~~~~~~~~~~~~~~~\nonumber\\
s_{5}=0,~~~~~~~S_{5}=-m-W(1+K^{-});\nonumber\\
s_{6}=0,~~~~~~~S_{6}=-m+W(1+K^{-});\nonumber\\
s_{7}=0,~~~~~~~~S_{7}=WK^{-};~~~~~~~~~~~~~~~~\nonumber\\
s_{8}=-\alpha,~~~~~S_{8}=-WK^{-}.~~~~~~~~~~~~~~
\end{eqnarray}
Here the following notations are introduced:  $K^{\pm}=-\alpha \Big((pu)\pm\sqrt{(up)^{2}-W^{2}}\Big)/W^{2}$,\   $W=\sqrt{p^{2}}$.

The solutions of the eigenvalue problem (\ref{eq:5}) in this case have the form:	
\begin{equation}
\lambda_{1,2}=-m \pm W\sqrt{1+2K^{+}}, \nonumber
\end{equation}
\begin{equation}\label{lamSM}
\lambda_{3,4}=-m \pm W\sqrt{1+2K^{-}},
\end{equation}
\begin{equation}
\Pi_{1,2}=\Sigma^{-}\cdot \frac{1}{2} \left[1\pm \hat{n}\ \frac{1+K^{+} - \gamma^{5}K^{+}}{\sqrt{1+2K^{+}}}  \right], \nonumber
\end{equation}
\begin{equation}\label{proSM}
\Pi_{3,4}=\Sigma^{+}\cdot \frac{1}{2} \left[1\pm \hat{n}\ \frac{1+K^{-} - \gamma^{5}K^{-}}{\sqrt{1+2K^{-}}}  \right].
\end{equation}

Vanishing of eigenvalues (\ref{lamSM}) gives the dispersion equation for neurino in moving matter
\begin{equation}\label{dispSM}
p^2-m^2 -2\alpha \left( (up) -s \sqrt{(up)^2 -p^2}\right) = 0 , 
\end{equation}
where $s=\pm 1$ is the spin projection on the axis $z$ (\ref{vec_z}).  If we restrict ourselves by the first order in $G_F$, then we have from (\ref{dispSM}) for solution with positive energy
\begin{equation}\label{dispSM+}
p^0 = \varepsilon + \frac{\alpha}{\varepsilon}\left[ (u^0\varepsilon - {\bf u}{\bf p}) -s \sqrt{(u^0\varepsilon - {\bf u}{\bf p})^2 - m^2}\right]  , 
\end{equation}
where $\varepsilon = \sqrt{{\bf p}^2 +m^2}$. Solution for negative energy is obtained by substitution $\varepsilon \to -\varepsilon$.  The dispersion law (\ref{dispSM}) coincides with equation obtained in \cite{Pivovarov:2005cu}, \cite{Studenikin:2008qk} at investigation of neutrino propagator in moving media.

In case of Standard Model it is easy to convince yourself that the spin projection on the axis of complete polarization is not a conservative value. The Hamiltonian is defined by Dirac operator (\ref{dddf})
\begin{equation}
H=p^{0}-\gz S.
\end{equation}
We can use a known zero commutator
\begin{equation}
[R,S]=0,~~~~R=\gf\hat{z}\hat{n},
\end{equation}
for simple calculation of commutator $R$ with Hamiltonian
\begin{equation}
[R,H]=\gz[S,R]+[\gz,R]S=[\gz,R]S,
\end{equation}
which may be reduced to $[\gz,R]$. With use of the standard representation of  $\gamma$-matrices we have
\begin{equation}
R=
\begin{pmatrix}
\bs\bf{v}&-i\bs\bxi\\
-i\bs\bxi&\bs{\bf{v}}
\end{pmatrix}, ~~{\bf{v}}=n^{0}{\bf{z}}-z^{0}{\bf{n}}, ~~\bxi=[{\bf{z}}\times {\bf{n}}].
\end{equation}

If to require  $[\gz,R]=0$, we come to condition $\bxi=0$, i.e.	
\begin{equation} \label{conse}
\bxi\equiv[{\bf{z}}\times {\bf{n}}]=bW[{\bf{p}}\times{\bf{u}}]=0.
\end{equation}
Thus, a spin projection on the axis $z^{\mu}$ is conserved only in the case ${\bf{u}}_{\perp}=0$,    
when 3-momentum of propagator coinsides in direction (or opposite) with matter velocity. In this case the found polarization vector $z^\mu$ (\ref{vec_z}) takes the form
\begin{equation} \label{heli}
z^{\mu}=\frac{1}{W} \left( | \mbox{\bf{p}} | ,\ p^{0}\frac{\bf{p}}{|\bf{p}|}  \right), 
\end{equation}
which corresponds to helicity state of fermion, but the off-shell one since $W\not= m$.

In general case, at arbitrary direction of matter velocity, in spite of   $[\Sigma^{\pm},S]=0$, the spin projection on the axis $z^{\mu}$ is not conserved: $[\Sigma^{\pm},H]\neq 0$. Evidently, that for propagator of more general form than the Standard Model one, the spin projection on the axis of complete polarization  $z$ also is not a conservative value.

\subsubsection{Rest matter case} 

Let us consider in detail a particular case of SM propagator (\ref{dddf}), when matter is in the rest  (${\bf{u}}=0, u_{0}=1$). In this case, according to Eq. (\ref{conse}), spin projection is concerved and polarization vector $z^\mu$  also corresponds to helicity state (\ref{heli}).

Straight calculation shows that the generalized spin projectors (\ref{Sigma}) in this case are projectors onto the spatial momentum direction
\begin{equation}
\Sigma^{\pm}=\frac{1}{2}\Big(1\pm\bS\frac{\bf{p}}{|\bf{p}|}\Big),~\bS=\gamma^{0}\bg\gamma^{5}.
\end{equation}

For the rest matter the eigenvalues and eigenprojectors are particular case of expressions  	(\ref{lamSM}), (\ref{proSM}) and look as follows:
\begin{equation}
\lambda_{1,2}=-m \pm W\sqrt{1-\frac{2\alpha}{W^{2}}(E+|\bf{p}|)},
\end{equation}
\begin{equation}
\lambda_{3,4}=-m \pm W\sqrt{1-\frac{2\alpha}{W^{2}}(E-|\bf{p}|)},
\end{equation}
\begin{equation}\label{vvvbgrtewsdf}
\Pi_{1}=\frac{1}{4}(1-\bS\frac{\bf{p}}{|\bf{p}|})\Big(1+\frac{\hat{n}}{B^{+}}[1-\frac{\alpha(E + |\bf{p}|)}{W^{2}}(1-\gamma^{5})]\Big),
\end{equation}
\begin{equation}
\Pi_{2}=\frac{1}{4}(1-\bS\frac{\bf{p}}{|\bf{p}|})\Big(1-\frac{\hat{n}}{B^{+}}[1-\frac{\alpha(E + |{\bf{p}}|)}{W^{2}}(1-\gamma^{5})]\Big),
\end{equation}
\begin{equation}
\Pi_{3}=\frac{1}{4}(1+\bS\frac{\bf{p}}{|\bf{p}|})\Big(1+\frac{\hat{n}}{B^{-}}[1-\frac{\alpha(E - |\bf{p}|)}{W^{2}}(1-\gamma^{5})]\Big),
\end{equation}
\begin{equation}
\Pi_{4}=\frac{1}{4}(1+\bS\frac{\bf{p}}{|\bf{p}|})\Big(1-\frac{\hat{n}}{B^{-}}[1-\frac{\alpha(E - |{\bf{p}}|)}{W^{2}}(1-\gamma^{5})]\Big),
\end{equation}
where $B^{\pm}=\sqrt{1-\frac{2\alpha}{W^{2}}(E \pm |\bf{p}|)}$.

Thus, for the rest matter the well-known fact \cite{Man87,Pan92} is reproduced that neutrino with definite helicity has a definite law of dispersion in matter.

If  some eigenvalue is vanished,  we obtain a dispersion relation -- energy and momentum connection. We have for $\lambda_{1,2}$
\begin{equation}
E^{2}-2\alpha E -m^{2}-{\bf{p}}^{2}-2\alpha |{\bf{p}}|=0,
\end{equation}
\begin{equation}
E_{1,2}=\alpha\pm\sqrt{(|{\bf{p}}|+\alpha)^{2}+m^{2}},
\end{equation}
and for $\lambda_{3,4}$:
\begin{equation}
E^{2}-2\alpha E -m^{2}-{\bf{p}}^{2}+2\alpha |{\bf{p}}|=0,
\end{equation}
\begin{equation}
E_{3,4}=\alpha\pm\sqrt{(|{\bf{p}}|-\alpha)^{2}+m^{2}},
\end{equation}

\section{The propagation of neutrino in an external magnetic field}

In previous section we found that in moving matter there exists an axis of complete polarization $z^{\mu}$ (\ref{vec_z}), and corresponding spin projectors (\ref{Sigma}) commute with the  propagator. A similar situation arises when  neutrino propagates in a magnetic field.

An inverse propagator  of a neutral fermion with an anomalous magnetic moment $ \mu $ in a constant external electromagnetic field is as follows:
\begin{equation}\label{proppp}
S=\hat{p}-m-\frac{i}{2}\mu\sigma^{\alpha\beta}F_{\alpha\beta}, \ \ \ \ \ \ \sigma^{\alpha\beta}=\frac{1}{2}[\gamma^{\alpha},\gamma^{\beta}],
\end{equation}
where $F_{\alpha\beta}$ is the electromagnetic field tensor. In the case of a magnetic field, it takes more customary form:
\begin{equation}\label{new_propagator}
S=\hat{p}-m+\mu\bS{\bf{B}},~~~~ \bS=\gz\bg\gf.
\end{equation}

Having electromagnetic field tensor and 4-momentum, we can construct a polarization vector $z^{\mu}$  ($ z^{2}=-1$ and $z_{\mu}p^ {\mu}=0$):
\begin{equation}\label{zB}
z^{\mu}=b\epsilon^{\mu\nu\lambda\rho}F_{\nu\lambda}p_{\rho},~~~~~~~ {b}=(p_{0}^{2}{\bf{B}}^{2}-({\bf{p}}{\bf{B}})^{2})^{-1/2}.
\end{equation}

Using this vector\footnote{This vector arises in consideration of motion of a charged relativistic fermion  in a constant and homogenious magnetic field, see 4-th edition of textbook \cite{AB}, \S 1.6 . We consider another situation: neutral fermion with an anomalous magnetic moment in a magnetic field, but it turns out that in this case the constructed spin projector also commutes with the propagator.} we can construct a spin projector with the same properties as in the case of neutrino propagation in a matter:

\begin{equation}\label{sb}
\Sigma^{\pm}=\frac{1}{2}(1 \pm \gamma^{5}\hat{z}).
\end{equation}

It is easy to see that the spin projectors commute with the inverse propagator (\ref{new_propagator}). To this end, let us write down the particular case of the vector $z^{\mu}$ in magnetic field
\begin{equation}\label{zB1}
z^{\mu}=b(({\bf{B}}{\bf{p}}),p^{0}{\bf{B}}),~~~~
b=(p_0^2{\bf B}^2 - ({\bf{B}}{\bf{p}})^2)^{-1/2}
\end{equation}
and matrix $\gf\hat{z}$ may be  rewritten as
\begin{equation}\label{R_equi}
R \equiv \gf\hat{z}=b(\gf\gz({\bf{B}}{\bf{p}})+p^{0}\gz(\bS{\bf{B}})),~~~~R^{2}=1.
\end{equation}
After this, it is easy to see that $[S,\Sigma^{\pm}]=0$.

Further we can apply the same trick that was used for the propagator in a matter: it was found that ``under observation'' of the spin projector, the gamma-matrix structures are simplified. So we can act by unit matrix composed of spin projectors onto the inverse propagator and to obtain two orthogonal terms $S^+$ and $S^-$: 
\begin{equation}
S=(\Sigma^{+}(z)+\Sigma^{-}(z))S \equiv S^+ + S^-.
\end{equation}

Since two matrices commute $[S,R]=0$, they have a common eigenvector: 
\begin{equation}
S\Psi = \lambda\Psi ,\ \ \ \gf\hat{z}\Psi =\sigma\Psi ,\ \ \ \sigma=\pm1.
\end{equation} 

The eigenvector of the operator $R$ is obvious: $\Psi^{\pm}=\Sigma^{\pm} \Psi_0$, therefore the system looks like this:
\begin{equation}
S^{\pm}\Psi^{\pm}=\lambda\Psi^{\pm}, \ \ \ \ \  \gf\hat{z}\Psi^{\pm}=\pm \Psi^{\pm} .
\end{equation} 

Since the eigenvalues of the matrix $R$ are equal to $\pm1$, from (\ref{R_equi}) we can find the useful relation 
\begin{equation}
(\bS{\bf{B}})\Psi^{\pm}=\frac{1}{p^{0}}(\gf({\bf{p}}{\bf{B}}) \pm \gz\frac{1}{b}) \Psi^{\pm}.
\end{equation}
Then, in analogy with the case of matter, in the $S^{\pm}$ contributions the $\gamma$-matrix structure can be transformed. Instead of (\ref{new_propagator}) we get
\begin{equation}\label{s_plus_minus}
S^{\pm}=\Sigma^{\pm}(z)\Big[\hat{p}-m+\frac{\mu}{p^{0}}(\gf({\bf{p}}{\bf{B}}) \pm \gz\frac{1}{b})\Big].
\end{equation}

Let us recall that for covariant matrix of the form
\begin{equation}\label{cov_m}
S=aI+b\hat{p}+c\gf+d\hat{p}\gf
\end{equation}
solutions of the matrix eigenvalue problem are known \cite{q12} and were used for the propagator in matter (\ref{eig_1}).

The inverse propagator in the external field (\ref{new_propagator}), (\ref{s_plus_minus}) is non-covariant (in particular, it contains $\gamma^0$), but for algebraic problem this is not so important. At solving of eigenvalue problem with the matrix (\ref{cov_m}), the momentum vector $p^{\mu}$ may be changed by any four numbers. Therefore, if we redefine the vector $p^{\mu}$ in $S^{\pm}$, we can get rid of $\gamma^0$ and use the ready answer for eigenvalues and eigenprojectors.  

So, we can introduce ``4-vector''
\begin{equation}
p_{\pm}^\mu= (p^{0} \pm \frac{\mu}{bp_{0}}, \ \bf{p} ) 
\end{equation} 
and after this, the inverse propagator takes the form:
\begin{equation}\label{obr_prop}
S^{\pm}=\hat{p}_{\pm}-m+\mu\gf\frac{({\bf{B}}{\bf{p}})}{p_{0}},
\end{equation}
in which there are only $I$, $\hat{p}_{\pm} $ and $\gf$ matrix, and which is algebraically similar to the vacuum propagator. Therefore, we can use the formulas (\ref{eig_1}) for eigenvalues and eigenprojectors:
\begin{eqnarray}\label{field}
\lambda_{1}^{\pm}=-m + \sqrt{W_{\pm }^{2}+\frac{\mu^{2}}{p_{0}^{2}}({\bf{B}}{\bf{p}})^{2}},\\
\lambda_{2}^{\pm}=-m - \sqrt{W_{\pm }^{2}+\frac{\mu^{2}}{p_{0}^{2}}({\bf{B}}{\bf{p}})^{2}},\\
\Pi_{1}^{\pm}=\frac{\Sigma^{\pm}}{2}\Big(1-\frac{1}{A^{\pm}}(\hat{p}_{\pm}+\frac{\mu ({\bf{B}}{\bf{p}})}{p_{0}}\gf)\Big),\\
\Pi_{2}^{\pm}=\frac{\Sigma^{\pm}}{2}\Big(1+\frac{1}{A^{\pm}}(\hat{p}_{\pm}+\frac{\mu ({\bf{B}}{\bf{p}})}{p_{0}}\gf)\Big).
\end{eqnarray}
We introduced here the notations: $W_{\pm}=\sqrt{p^{2}_{\pm}}$, \ \  $A^{\pm}=\sqrt{W_{\pm }^{2}+\mu^{2} ({\bf{B}}{\bf{p}})^{2}/p_{0}^{2}}$.  If the eigenvalue is vanishing, we can obtain the well-known dispersion law for movement of anomalous magnetic moment in magnetic field \cite{TBKh,Bag90}
\begin{equation}
E^2 = m^2+ {\bf{p}}^2 + \mu^2 {\bf{B}}^2 \pm 2\mu \sqrt{m^2 {\bf{B}}^2 + {\bf{p}}^2 {\bf{B}}^2_\perp}.
\end{equation}
Here $\pm$ corresponds to different signs in \eqref{obr_prop}, i.e. to terms $S^\pm$ in propagator, which are accompanied by spin projectors   $\varSigma^{\pm}$.

The spectral representation of the inverse propagator with found eigenvalues and eigenprojectors   can be written as:
\begin{equation}
S=\sum_{i=1}^{2}\lambda_{i}^{+}\Pi_{i}^{+}+\sum_{i=1}^{2}\lambda_{i}^{-}\Pi_{i}^{-}.
\end{equation}

So, in constant magnetic field all eigenvalues are classified by the spin projection on the fixed axis $z$ (\ref{zB1}). It turns out that, as in the case of moving medium, the projection on this axis, in general, is not a conserved quantity.

The inverse propagator (\ref{new_propagator}) may be connected with the Dirac Hamiltonian
\begin{equation}
S=\gamma^0 (p^0 - H_D),\ \ \ \ \  H_D= {\bm \alpha}{\bf p} + \beta m + \mu \gamma^0 (\bf{\Sigma} \bf{B}) .
\end{equation}
Using the zero commutator of the matrix $R=\gamma^5\hat{z}$ with the inverse propagator
\begin{equation}
0 = [R,S] = \gamma^0 [H_D, R] + [R, \gamma^0] (p^0 - H_D) ,
\end{equation}
we can reduce the case to the commutator $[R, \gamma^0]$.  Calculating it in the standard representation of gamma-matrices, we have
\begin{equation}
[\gamma^5 \hat{z}, \gamma^0] =   
\begin{pmatrix}
0 &  2z^0\\
2z^0 & 0
\end{pmatrix},\ \ \ \ \ z^0=b\ ({\bf{B}}{\bf{p}}).
\end{equation}
So we see that the projection of the spin on the axis of complete polarization (\ref{zB}) is conserved only in case of a transverse magnetic field.

\section{Conclusions}

In the present paper we have built the spectral representation of neutrino propagator both in a moving matter and in a constant external magnetic field. In this form (\ref{RS}), (\ref{spec_G}), which is based on the eigenvalue problem for inverse propagator (\ref{eq:5}), the propagator looks like a sum of poles accompanied by own $\gamma$-matrix orthogonal projectors. The advantage of this representation is that a single term in this sum is related only with one dispersion law for particle in media. More exactly, relation of energy and momentum appears as a result of vanishing of one of eigenvalues  
$\lambda_i=0$ in (\ref{spec_G}).

It turned out that both in matter and in magnetic field there exists the fixed 4-axis of complete polarization $z^\mu$, such that  all eigenvalues of propagator (and, consequently, dispersion laws)	are classified accoding to spin projection on this axis.  The found generalized spin projectors (\ref{Sigma}), (\ref{sb}) on the axis of complete polarization  play a special role in the eigenvalue problem, simplifying essentially algebraic calculations.

In the case of moving matter the states with the definite spin projection on the found axis  (\ref{vec_z}) have a definite dispersion law. In particular case of rest matter the operators  $\Sigma^{\pm}$ are projectors on  the helicity states in correspondence with known earlier results  \cite{Man87,Pan92}. Let us  emphasize that for moving matter or magnetic field the vanishing of commutator with inverse propagator $S$ $[S,\Sigma^{\pm}]=0$ does not lead to conservation of spin projection on this axis, since the spin projectors  $\Sigma^{\pm}$ do not commute, generally speaking, with Hamiltonian.

Let us note that the propagator in external magnetic field (after use of the $\Sigma^{\pm}$ properties) is not covariant one, it contains also $\gamma^0$ matrix besides the unit matrix, $\hat{p}$ and $\gamma^5$.	But for the eigenvalue problem the covariance is not essential, so after transfer to non-covariant ``momentum''   $p^\mu_\pm$ one can use an algebraic construction for vacuum propagator.

We considered here the cases of moving non-polarized matter or external magnetic field. In our approach one can take into account the matter polarization: it leads to simple substitution of four-velocity of matter  $u^{\mu}$ by some combination of velocity and matter polarization (see \cite{Lobanov:2001ar}), after it the same algebraic construction is repeated.

The most evident development of this approach is related with neurtino oscillation in matter, in particular, in astrophysical problems, for instance, in propagation of neutrinos through supernova envelope, see e.g. \cite{Duan:2009cd}, \cite{zas1}. The presence of the fixed axis of complete polarization and reducing the number of gamma-matrix structures should make this problem  algebraically similar to the mixing problem in vacuum -- see corresponding spectral representation in  \cite{q98}.  We suppose that dynamics of the neutrino spin in media in the presence of the off-shell axis of complete polarization also may be interesting.

We are grateful to V.A. Naumov, S.E. Korenblit, S.I. Sinegovsky and  A.V.Sinitskaya for useful discussions and comments.

%
%





\appendix
\section{Spectral representation of matrix of general form}  \label{sec:eigen}

In order to build a spectral representation of the matrix $S$ of general form, one needs to solve two eigenvalue problems.

Left eigenvalue problem:
\begin{equation}\label{LL}
S \psi =\lambda \psi
\end{equation}

and right one:
\begin{equation}\label{RR}
\phi^T S  = \phi^T \lambda. 
\end{equation}
Here $S$  is matrix of dimension $n$ and $\psi$, $\phi$ are the columns of this dimension.

Let us indicate the main properties of these problems.
\begin{itemize}
	\item The spectra of the left and right problems coinsides. Indeed, the eigenvalues of the left problem are defined by equation $det(S-\lambda E) = 0$, as for spectrum of the right -- it is defined by transpose matrix  $det(S-\lambda E)^T = 0$.
	\item Orthogonality of eigenvectors. Let us write down two equations 
	\begin{equation}\label{LL1}
	S \psi_i =\lambda_i \psi_i. 
	\end{equation}
	\begin{equation}\label{RR1}
	\phi_k^T S  = \phi_k^T \lambda_k. 
	\end{equation}
	Let us multiply (\ref{LL1}) by $\phi_k^T$  from the left, (\ref{RR1}) by  $\psi_i$ from the right and subtruct one equation from another. We have
	\begin{equation}\label{}
	0  = (\lambda_i - \lambda_k) \phi_k^T \psi_i , 
	\end{equation}
	i.e. eigenvectors of left and right problems $\phi_k$, $\psi_i$ are orthogonal \footnote{Case of degenerate eigenvalues -- see below an example of spectral representation of propagator.}  at  $i \not= k$.
	\begin{equation}\label{}
	\phi_k^T \psi_i  =  \psi_i^T  \phi_k \equiv (\psi_i, \phi_k)=0 \ \ \mbox{at} \ \ i \not= k
	\end{equation}
	One can require the orthonormality of these two sets of vectors
	\begin{equation}\label{orto}
	(\psi_i, \phi_k)=\delta_{ik}.
	\end{equation}
	\item Having solutions of both left and right problems with the property (\ref{orto}), one can build matrices of the form 
	\begin{equation}\label{}
	\Pi_i =  \psi_i  \phi_i^T ,\ \ \ \ \ i=1\dots n ,
	\end{equation}
	which are the set of orthogonal projectors.
	\begin{equation}\label{}
	\Pi_i \Pi_k = \delta_{ik} \Pi_k 
	\end{equation}
	Note that the projectors $\Pi_i$ (eigenprojectors) are the matrix solution of both left and right eigenvalue problems.
	\item  In particular case of hermitian matrix $S$, solutions of left and right problems are related as follows 
	\begin{equation}\label{}
	\phi_i =  \psi_i^* 
	\end{equation}       
	and eigenprojectors look like: 
	\begin{equation}\label{}
	\Pi_i =  \psi_i  \psi_i^\dagger ,\ \ \ \ \ i=1\dots n .
	\end{equation}
\end{itemize}
Having solutions of left and right problems, one can represent matrix in a form
\begin{equation}\label{}
S =  \sum_{i=1}^{n} \lambda_i \Pi_i = \sum_{i=1}^{n} \lambda_i \psi_i  \phi_i^T .
\end{equation} 
This is a spectral representation of a general form matrix, which includes solutions of both left $\psi_i $ and right $\phi_i $ eigenvalue problem.  

\section{Spectral representation of fermion propagator in vacuum and media} 
\label{sec:prop}

Let us consider the dressed fermion propagator in vacuum $S(p)=\hat{p}-m - \Sigma(p)$. In case of P-parity violating the self-energy $\Sigma(p)$ contains $\gamma^5$ and may be written as
\begin{equation}\label{}
\Sigma(p) =  a(p^2) + b(p^2)\hat{p}+c(p^2)\gamma^5 + d(p^2) \hat{p}\gamma^5  .
\end{equation}  
Note that in the absence of $\gamma^5$ a spectral representation is rather evident, but parity violation generates non-trivial eigenprojectors. Solving the eigenvalue problem for such matrix $S(p)$ of dimension 4, we will find only two eigenvalues  $\lambda_i(p^2)$ and eigenprojectors $\Pi_i(p)$.
Evidently,  this degeneration is related with spin and complete set of eigenprojectors is constructed with use of spin projectors, dependent on arbitrary polarization four-vector $s$.
\begin{equation}\label{}
\Sigma^{\pm} (s) \Pi_i(p), \ \ \ \ \ \ i=1,2  .
\end{equation}   
The appeared here generalized spin progectors \cite{q98} commute with propagator and in theories with parity violation take the form
\begin{equation}\label{}
\Sigma^{\pm} (s) =\frac{1}{2}(1 \pm \gamma^5 \hat{s} \hat{n}) ,\ \ \ \ \ \ n^\mu=p^\mu/W,\ \ \ \ \ \ W=\sqrt{p^2} .
\end{equation} 
The additional factor $\hat{n}$ in spin projector plays an essentional role in the dressed propagator, for bare propagator (and in theory without $\gamma^5$) it turnes into unit matrix.

Spectral representation for dressed inverse propagator in vacuum may be written in the following elegant form
\begin{eqnarray}\label{sp_vac}
S(p) = \lambda_1 \Sigma^{+} (s)  \Pi_1(p) + \lambda_1 \Sigma^{-} (s)  \Pi_1(p) +\nonumber\\+ \lambda_2 \Sigma^{+} (s)  \Pi_2(p) +\lambda_2 \Sigma^{-} (s)  \Pi_2(p) = \nonumber\\
= \lambda_1 \Pi_1(p) +  \lambda_2  \Pi_2(p) .
\end{eqnarray} 
It should be mentioned that for fermion propagator it is convenient to solve an eigenvalue  problem in a matrix form (i.e. to look for eigenprojectors insteed of eigenvectors), using the  $\gamma$-matrix basis. In doing so we will find only two eigenvalues for dressed propagator. As for eigenprogectors, the left problem has two matrix solutions  $\Pi_1, \Pi_2$ and there is no necessity to solve a right eigenvalue problem --- the orthogonality requirement fixes ambiguity completely  \cite{q12}. 

For propagator in a moving matter  (see (\ref{formula2})) there appear new  $\gamma$-matrix structures and it leads to disappearing of degeneration: in this case we have four different eigenvalues and eigenprojectors (\ref{eig_1}),  (\ref{eig_2}). Spectral representation of inverse propagator in matter 
\begin{eqnarray}\label{}
S(p,u)= \sum_{i=1}^{4} \lambda_i \pi_i(p,u) = \nonumber \\= \sum_{1}^{2} \lambda_i \Sigma^{-}(z) \Pi_i +  \sum_{3}^{4} \lambda_i \Sigma^{+}(z) \Pi_i 
\end{eqnarray} 
contains  the generalized spin projectors, depending of the fixed (in contrast to (\ref{sp_vac})) four-vector $z$ (\ref{vec_z}).

\end{document}